\documentclass[journal]{IEEEtran}
\usepackage{cite}
\usepackage{graphicx}
\usepackage{nohyperref} % disables hyperref, but allows syntax to remain without causing errors %%% as required for IEEE ComSoc submissions
\usepackage{amsmath}
\usepackage{amssymb}
\usepackage{array}
\usepackage{url}
\usepackage{color}

\begin{document}
\title{Path Loss in Reconfigurable Intelligent Surface-Enabled Channels}
\author{S.W.~Ellingson,~\IEEEmembership{Senior Member,~IEEE}% <-this % stops a space
\thanks{S.W.~Ellingson is with the Dept.\ of Electrical and Computer Engineering, Virginia Tech, Blacksburg, VA, 24061 USA e-mail: ellingson.1@vt.edu.}% <-this % stops a space
}
\maketitle
\thispagestyle{empty} %%% %%% Remove page numbers as required for IEEE ComSoc submissions

\begin{abstract}
A reconfigurable intelligent surface (RIS) employs an array of individually-controllable elements to scatter incident signals in a desirable way; for example, to facilitate links between base stations and mobile stations that would otherwise be blocked. A principal consideration in the study of RIS-enabled propagation channels is path loss. This paper presents a simple yet broadly-applicable method for calculating the path loss of a channel consisting of a passive reflectarray-type RIS. This model is then used to characterize path loss as a function of RIS size, link geometry, and the method used to set the element states. Whereas previous work presumes either (1) an array of parameterizable element patterns and spacings (most useful for analysis of specific designs) or (2) a continuous electromagnetic surface (most useful for determining scaling laws and theoretical limits), this work begins with (1) and is then shown to be consistent with (2), making it possible to identify specific practical designs and scenarios that exhibit the performance predicted using (2). This model is used to further elucidate the matter of path loss of the RIS-enabled channel relative to that of the free space direct and specular reflection channels, which is an important consideration in the design of networks employing RIS technology.
\end{abstract}

% \begin{IEEEkeywords}
% Reconfigurable intelligent surfaces, electromagnetic propagation, path loss, antenna arrays, antenna radiation patterns, beamforming
% \end{IEEEkeywords}

% \fi
%\IEEEpeerreviewmaketitle

\section{Introduction}
\label{sIntro}

%\IEEEPARstart{T}{his} demo file is intended to serve as a ``starter file''.
A reconfigurable intelligent surface (RIS) is a device that scatters signals in a controlled manner in order to improve 
% path loss, fading, and/or other characteristics of 
the propagation channel between transmitters and receivers; see e.g.\ \cite{DiRenzo_202011} and references therein.
RISs are often envisioned to be some form of reflectarray in which control consists of changing the phase and possibly magnitude of the electromagnetic field scattered by each element individually; this class of RISs is the focus of this paper.
An important consideration in RIS engineering is the path loss in the transmitter-RIS-receiver channel as a function of RIS size, link geometry, and element states. 

Analysis of path loss requires a method for calculating scattering by the RIS, for which there are two general strategies.
In the first strategy, scattering is computed as the discrete sum of fields scattered from elements; see e.g. \cite{Tang+9_1911}.
This is ideal for practical analysis and design problems since the relationship between path loss and the element pattern and spacing within the RIS is explicit.
The principal disadvantage of this strategy is that \emph{embedded} element pattern (i.e., the pattern of the element accounting for mutual coupling) is difficult to calculate.
This issue is commonly bypassed by assuming that the responses of the elements are identical and easily parameterizable, which is known to be a reasonable approximation for large planar arrays of regularly-spaced elements \cite{ST13}.  

In the second strategy, the RIS is treated as an continuous surface subject to electromagnetic boundary conditions (typically surface impedance or surface currents) which, when combined with Maxwell's equations, yields a boundary value problem that can be solved for the scattered fields; see e.g., \cite{Dardari_202011,BS20,NJSP20}.
This is ideal for identifying theoretical limits of performance and useful ``scaling laws'' (e.g., how path loss varies with RIS area, path length, and so on) that are independent of the specific technology used to implement the RIS.
However this strategy is not particularly useful for the analysis and design of specific RIS devices because the relationship between the boundary conditions and the element pattern(s), spacing, and element states comprising a specific RIS is not explicit and may be difficult to ascertain.
%; typically requiring the use of Maxwell's equations and a full-wave solver. 

The principal contribution of this work is a method for calculation of path loss following the first strategy (i.e., beginning with an array of parameterizable element patterns and spacings) which is similar to \cite{Tang+9_1911}, and which is then used to derive a particular design for which the calculated path loss is equal to that predicted by the second strategy in the special case where the boundary condition corresponds a perfectly conducting plate having area equal to the RIS. 
This particular design consists of elements having directivity of about 5~dBi with half-wavelength spacing, which essentially confirms that the results obtained using the second strategy are applicable to a typical practical reflectarray-type RIS.

A second contribution of this work is the use of this model to further elucidate the matter of path loss of the transmitter-RIS-receiver channel relative to that of the free space transmitter-receiver and specular reflection channels, which has been a source of confusion in the literature.  While this issue has been previously addressed in \cite{BS20,NJSP20,Ozdogan+2_1910,GSK20,BOL20}, the analysis in this paper yields a concise independent summary which may be useful in better understanding this topic.
 
%%%%%%%%%%%%%%%%%%%%%%%%%%%%%%%%%%%%%%%%%%%%%%%%%%%%%%%%%%%%%%%%%%%%%%%%%%%%%%
\section{Received Power in the RIS-Enabled Channel}
%%%%%%%%%%%%%%%%%%%%%%%%%%%%%%%%%%%%%%%%%%%%%%%%%%%%%%%%%%%%%%%%%%%%%%%%%%%%%%
\label{sLPE}

\begin{figure} %[!t]
\begin{center}
\includegraphics[width=2in]{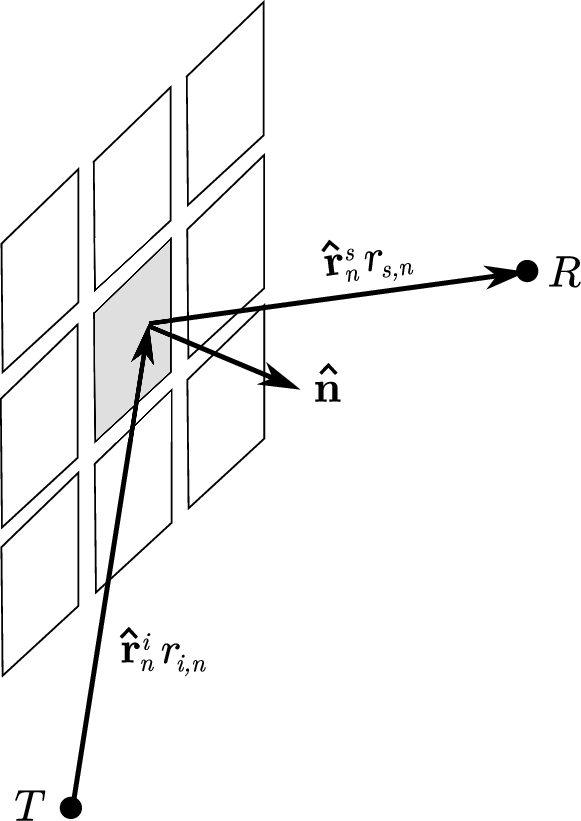}
\end{center}
\caption{Geometry for scattering from element $n$ of an $N$-element RIS.
The transmitter and receiver are indicated as $T$ and $R$ respectively.  
The unit vectors $\hat{\bf r}^i_n$ and $\hat{\bf r}^s_n$ are used to indicate directions from the transmitter (``incident'') and to the receiver (``scattered''), respectively.
The unit vector $\hat{\bf n}$ indicates the outward-facing perpendicular (``broadside'') direction.
}
\label{fGeo1}
\end{figure}

Referring to Fig.~\ref{fGeo1}, 
the spatial power density  
incident on the $n^{\mbox{th}}$ element is:
\begin{equation}
S^i_n = P_T G_T(\hat{\bf r}^i_n) / 4\pi r_{i,n}^2
\end{equation}
where
$P_T$ is the total power applied by the transmitter to the transmit antenna system,
$G_T(\hat{\bf r}^i_n)$ is the gain of the transmit antenna system in the direction $\hat{\bf r}^i_n$, and 
$r_{i,n}$ is the distance from the transmitter to the $n^{\mbox{th}}$ element.
The power captured by the $n^{\mbox{th}}$ element is
\begin{equation}
P^i_n =  S^i_n A_e(-\hat{\bf r}^i_n)
\end{equation}
where $A_e(-\hat{\bf r}^i_n)$ is the effective aperture of the element in the direction of the transmitter.
This is related to the gain $G_e(-\hat{\bf r}^i_n)$ of the element by 
$ A_e(-\hat{\bf r}^i_n) = G_e(-\hat{\bf r}^i_n) \lambda^2 / 4\pi $
%%
%\begin{equation}
%A_e(-\hat{\bf r}^i_n) = \frac{\lambda^2}{4\pi} G_e(-\hat{\bf r}^i_n) 
%\end{equation}
%%
where $\lambda$ is wavelength.
Thus:
\begin{equation}
%P^i_n =  P_T G_T(\hat{\bf r}^i_n) G_e(-\hat{\bf r}^i_n) \left(\frac{\lambda}{4\pi r_{i,n}}\right)^2  
P^i_n =  P_T G_T(\hat{\bf r}^i_n) G_e(-\hat{\bf r}^i_n) \left(\lambda / 4\pi r_{i,n}\right)^2  
\end{equation}
The power density at the receiver due to scattering from the $n^{\mbox{th}}$ element is:
\begin{equation}
S^s_n = P^s_n G_e(\hat{\bf r}^s_n) / 4\pi r_{s,n}^2
\end{equation}
where
$P^s_n$ is the power applied by the RIS to the element, 
$\hat{\bf r}^s_n$ is the direction from the element to the receiver, and
$r_{s,n}$ is the distance from the element to the receiver.
Note that the same pattern $G_e(\cdot)$ is used for both incidence and scattering, which is valid as long as the maximum dimension of the element is $\ll \lambda$.
The power captured by the receiver from the RIS element is
\begin{equation}
%P_{R,n} =  S^s_n \left[ \frac{\lambda^2}{4\pi}G_R(-\hat{\bf r}^s_n) \right]
P_{R,n} =  S^s_n \left[ G_R(-\hat{\bf r}^s_n) \lambda^2 / 4\pi \right]
\end{equation}
where 
$G_R(-\hat{\bf r}^s_n)$ is the gain of the receiver's antenna system in the direction of the element.

The efficiency $\epsilon_p=P^s_n/P^i_n$ 
accounts for the limited efficiency of practical antenna elements and insertion losses associated with components required to implement the desired change in magnitude and phase.  If the RIS is passive, $\epsilon_p\le 1$.  
Combining expressions, we obtain:
\begin{equation}
P_{R,n} = P_T  G_T(\hat{\bf r}^i_n) G_R(-\hat{\bf r}^s_n) \left(\frac{\lambda}{4\pi}\right)^4\frac{G_e(-\hat{\bf r}^i_n)G_e(\hat{\bf r}^s_n)}{r_{i,n}^2 r_{s,n}^2} ~\epsilon_p
\label{eLE1}
\end{equation}
The voltage-like signal observed by the receiver is:
\begin{equation}
y = \sum_{n=1}^N b_n \sqrt{P_{R,n}} ~e^{j\phi_n}   
\end{equation}
where 
\begin{equation}
\phi_n = \left( r_{i,n}+r_{s,n}\right) 2\pi / \lambda
\end{equation}
is the phase accrued by propagation over the path that includes the $n^{\mbox{th}}$ element, and
the $b_n$'s are complex-valued coefficients representing the controlled responses of the elements.
%The power $P_R$ observed at the receiver is maximized by setting the phase of $b_n$ equal to $-\phi_n$, which corresponds to coherent combining of the signals arriving along the $N$ paths at the receiver.  However this of course is not the only option.
%Finally, 
The total power $P_R$ observed by the receiver is therefore:
\begin{equation}
P_R = \left| \sum_{n=1}^N b_n \sqrt{P_{R,n}} ~e^{j\phi_n} \right|^2
\label{ePRsum}    
\end{equation}

%%%%%%%%%%%%%%%%%%%%%%%%%%%%%%%%%%%%%%%%%%%%%%%%%%%%%%%%%%%%%%%%%%%%%%%%%%%%%%
\section{Path Loss and Element Pattern}
%%%%%%%%%%%%%%%%%%%%%%%%%%%%%%%%%%%%%%%%%%%%%%%%%%%%%%%%%%%%%%%%%%%%%%%%%%%%%%
\label{sSPL}

%-----------------------------------------------------------------------------
\subsection{Path Loss} 
%-----------------------------------------------------------------------------

%It is often desirable to separate factors associated with the propagation channel (including the RIS) -- collectively referred to as \emph{path loss} -- from the antenna gains of the transmitter and receiver.
%To accomplish this, we make the following approximations:
In order to separate factors associated with the propagation channel (including the RIS) from the antenna gains of the transmitter and receiver, we make the following approximations:
(1) $G_T(\hat{\bf r}^i_n)$ is constant with respect to $n$; i.e., the gain of the transmitter is constant over the RIS.
(2) $G_R(-\hat{\bf r}^s_n)$ is constant with respect to $n$; i.e., the gain of the receiver is constant over the RIS.
%These approximations allow $G_T(\hat{\bf r}^i_n)$ and $G_R(-\hat{\bf r}^s_n)$ to be extracted from the sum in Equation~\ref{ePRsum}, yielding
Then:
\begin{flalign}
%P_R = &P_T G_T G_R \left(\frac{\lambda}{4\pi}\right)^4 \nonumber \\
%      &\cdot \left| \sum_{n=1}^N b_n \sqrt{\frac{G_e(-\hat{\bf r}^i_n)G_e(\hat{\bf r}^s_n)}{r_{i,n}^2 r_{s,n}^2}} ~e^{j\phi_n}  %\right|^2 \epsilon_p     
P_R = &P_T G_T G_R \left(\lambda / 4\pi \right)^4 \nonumber \\
      &\cdot \left| \sum_{n=1}^N b_n \sqrt{\frac{G_e(-\hat{\bf r}^i_n)G_e(\hat{\bf r}^s_n)}{r_{i,n}^2 r_{s,n}^2}} ~e^{j\phi_n}  \right|^2 \epsilon_p      
\label{eSPL_PR1}
\end{flalign}
This expression is exact if the transmit and receive antenna systems exhibit isotropic gain.
However, the approximation is broadly applicable.
In particular, this approximation is suitable for transmit and receive antenna radiation patterns which are only weakly directional, as is the case for mobile stations.
This approximation is also suitable for transmit and receive antenna systems exhibiting approximately constant gain over the angular span corresponding to the RIS -- this is possible even if the transmit and receive antenna systems form narrow beams, as long as the RIS is sufficiently far away. %, even if the RIS is not in the far field. 

Equation~\ref{eSPL_PR1} is in the form of the Friis transmission equation; therefore the path loss $L_{RIS}$ is given by:  
\begin{equation}
L_{RIS}^{-1} = \left(\frac{\lambda}{4\pi}\right)^4 
      \left| \sum_{n=1}^N b_n \sqrt{\frac{G_e(-\hat{\bf r}^i_n)G_e(\hat{\bf r}^s_n)}{r_{i,n}^2 r_{s,n}^2}} ~e^{j\phi_n} \right|^2 \epsilon_p             
\label{ePG1}
\end{equation}

%-----------------------------------------------------------------------------
\subsection{Element Pattern Model} 
\label{sGEPM}
%-----------------------------------------------------------------------------

Next, we seek a expression for the element radiation pattern $G_e(\cdot)$ that is simple yet broadly-applicable.
Since RIS elements are commonly envisioned to be electrically-small low-gain elements above a conducting ground screen, we choose the popular model (see e.g., \cite{ST13}, Sec.~9.7.3):
\begin{flalign}
G_e(\psi) &= \gamma\cos^{2q}(\psi) & 0\le\psi<\pi/2 \label{eGep} \\
          &= 0                     & \pi/2\le\psi\le\pi \label{eGep0}
\end{flalign} 
where 
$\psi$ is the angle measured from RIS broadside, 
$q$ determines the gain of the element, and 
$\gamma$ is the coefficient required to satisfy conservation of power.
Power is conserved by requiring the integral of $G_e(\psi)$ over a surface enclosing the element to be equal to $4\pi$~sr.
%%
%\begin{equation}
%\oint G_e(\psi)~d\Omega = 4\pi
%\end{equation}
%
It is shown in \cite{ST13} (Sec.~9.7.3) that this constraint is satisfied for
\begin{equation}
\gamma=2(2q+1)
\label{egamma}   
\end{equation}

An appropriate value of $q$ is determined from the broadside gain $G_e(\psi=0)$ of the element.
From Equations~\ref{eGep}--\ref{egamma}:
\begin{equation}
%q = \frac{1}{4}G_e(\psi=0) - \frac{1}{2}
q = G_e(\psi=0)/4 - 1/2
\end{equation}

For general studies of RIS-based wireless communications, it is awkward to choose $q$ (hence the element gain) to correspond to a particular element design.
Instead, a physically-motivated ``benchmark'' value of $q$ is preferred.
Such a value may be obtained by requiring $A_e(\psi=0)$ to be equal to $(\lambda/2)^2$. 
Under this condition, the physical area of a RIS is equal to the sum of the effective apertures of the elements when the elements are separated by $\lambda/2$.
This criterion is of particular interest because it essentially corresponds to the ``electromagnetic surface'' paradigm described as the ``second strategy'' in Section~\ref{sIntro}.  
Invoking this criterion, we require:
\begin{equation}
% A_e(\psi=0) = \frac{\lambda^2}{4\pi}G_e(\psi=0) = \left(\frac{\lambda}{2}\right)^2
A_e(\psi=0) = \left(\lambda^2/4\pi\right) G_e(\psi=0) = \left(\lambda/2\right)^2
\end{equation}
This yields $\gamma=\pi$, $q\cong 0.285$, and subsequently $G_e(\psi=0)\cong 5$~dBi. 
This value is consistent with the gain of typical patch antenna elements, which range between 3~dBi and 9~dBi (see e.g. \cite{ST13}, Sec.~11.2). 
We therefore define the desired benchmark value of $q$ to be $q_0 = 0.285$.
%%
%\begin{equation}
%q_0 = 0.285 ~~\mbox{(definition)}
%\end{equation}

At this point, it should be emphasized that this pattern model does not \emph{require} that the elements be spaced by $\lambda/2$; however if the elements are spaced in this manner, the choice of $q=q_0$ will result in the broadside effective aperture of the RIS being equal to its physical aperture.  
Similarly, the choice of $q=q_0$ is not required, and in fact this parameter can be ``tuned'' to model other specific element designs.  
%The recommended procedure if a different element gain is desired is to determine first the value of $\gamma$ that yields the desired broadside gain, and then to solve for $q$ using Equation~\ref{egamma}. 

In subsequent work, we shall assume that all element patterns are identical; i.e., constant with respect to $n$.  
In practice, the patterns of elements close to the center of a large RIS will be nearly symmetric and uniform, whereas the patterns of elements near the edge will exhibit some degree of asymmetry and gain variation \cite{ST13}.
Since the ratio of ``edge elements'' to ``interior elements'' is small for an electrically-large RIS, this variation will typically not significantly affect path loss calculation.
%Therefore, it is henceforth assumed that element patterns are identical; i.e., constant with respect to $n$.

%-----------------------------------------------------------------------------
\subsection{Alternative Form of the Path Loss Equation} 
%-----------------------------------------------------------------------------

A useful alternative form of Equation~\ref{ePG1} may be obtained using the element pattern model proposed in the previous section.
First, note that
$ \cos\psi = \hat{\bf r}(\psi)\cdot\hat{\bf n} $
%%
%\begin{equation}
%\cos\psi = \hat{\bf r}(\psi)\cdot\hat{\bf n}
%\end{equation}
%%
where 
$\hat{\bf r}(\psi)$ is a unit vector pointing outward from the element, indicating the direction in which the pattern is being evaluated; 
$\hat{\bf n}$ is the unit normal vector indicating RIS broadside, i.e., the direction corresponding to $\psi=0$; and 
``$\cdot$'' denotes the scalar (``dot'') product.
%\footnote{The principal motivation in employing the scalar product notation is that this greatly simplifies computation. This is because $\hat{\bf n}$ is known in advance and does not change, $\hat{\bf r}$ is easy to compute, and the scalar product requires only algebraic operations. In contrast, the general expression for $\psi$ is relatively complicated, requires trigonometric operations, and is a greater burden to compute. This additional burden can be significant when computing large amounts of performance data for a large RIS.}
Now assuming $q=q_0$, we find:
\begin{flalign}
G_e(-\hat{\bf r}^i_n) &= \pi \left(-\hat{\bf r}^i_n\cdot\hat{\bf n}\right)^{2q_0} ~\mbox{, and}\\
G_e(+\hat{\bf r}^s_n) &= \pi \left(+\hat{\bf r}^s_n\cdot\hat{\bf n}\right)^{2q_0}
\end{flalign} 
Thus, Equation~\ref{ePG1} becomes:
\begin{equation}
L_{RIS}^{-1} = \frac{\lambda^4}{256\pi^2}
      \left| \sum_{n=1}^N b_n \sqrt{\frac{\left(-\hat{\bf r}^i_n\cdot\hat{\bf n}\right)^{2q_0} \left(+\hat{\bf r}^s_n\cdot\hat{\bf n}\right)^{2q_0}}{r_{i,n}^2 r_{s,n}^2}} ~e^{j\phi_n} \right|^2 \epsilon_p
\label{ePG2}
\end{equation}
%
%Note that this expression does \emph{not} assume that the element spacing is $\lambda/2$; all that is assumed is that the element patterns are well-modeled by the ``standard'' pattern described at the end of the previous section.

%%%%%%%%%%%%%%%%%%%%%%%%%%%%%%%%%%%%%%%%%%%%%%%%%%%%%%%%%%%%%%%%%%%%%%%%%%%%%%
\section{Far Case}
%%%%%%%%%%%%%%%%%%%%%%%%%%%%%%%%%%%%%%%%%%%%%%%%%%%%%%%%%%%%%%%%%%%%%%%%%%%%%%
\label{sFFC}

In this section we consider the ``far'' case.
For the purposes of this paper, the RIS is said to be far from the transmitter if $\hat{\bf r}^i_n$ and $r_{i,n}$ are approximately independent of $n$; i.e., approximately equal to the same constants $\hat{\bf r}_i$ and $r_{i}$, respectively.
Similarly, the RIS is said to be far from the receiver if $\hat{\bf r}^s_n$ and $r_{s,n}$ are approximately equal to the same constants $\hat{\bf r}_s$ and $r_{s}$, respectively.
No approximation is made for the phases $\phi_n$: These values continue to be exact and are not assumed to be independent of $n$.

%-----------------------------------------------------------------------------
\subsection{Expressions for Path Loss in the Far Case} 
%-----------------------------------------------------------------------------

Under the far approximation, Equation~\ref{ePG2} simplifies to: 
\begin{equation}
L_{RIS}^{-1} = \frac{\lambda^4}{256\pi^2} \frac{\left(-\hat{\bf r}_i\cdot\hat{\bf n}\right)^{2q_0} \left(+\hat{\bf r}_s\cdot\hat{\bf n}\right)^{2q_0}}{r_i^2 r_s^2}
\left| \sum_{n=1}^N b_n e^{j\phi_n} \right|^2
\epsilon_p
\label{ePG3a}
\end{equation}
%
% Note that no approximation has been imposed on the path phases ($\phi_n$'s).
Equation~\ref{ePG3a} correctly indicates that path loss for the RIS in the far case is proportional to $r_i^2 r_s^2$, regardless of the chosen coefficients ($b_n$'s).
Equation~\ref{ePG3a} further indicates that path loss is minimized when the phase of $b_n$ is set equal to $-\phi_n$.
Assuming phase-only control of the elements, one would select $b_n=e^{-j\phi_n}$. 
In this case Equation~\ref{ePG3a} reduces to:
\begin{equation}
L_{RIS}^{-1} = \frac{\lambda^4}{256\pi^2} N^2 \frac{\left(-\hat{\bf r}_i\cdot\hat{\bf n}\right)^{2q_0} \left(+\hat{\bf r}_s\cdot\hat{\bf n}\right)^{2q_0}}{r_i^2 r_s^2} \epsilon_p
\label{ePG3}   
\end{equation}
Recall that the $q=q_0$ element pattern proposed in Section~\ref{sGEPM} was derived from the constraint that the sum of the broadside effective apertures of the elements is equal to the physical area $A$ of the RIS when the element spacing is equal to $\lambda/2$.  If we now commit to this element spacing, then
$ A = N \left(\lambda/2\right)^2 $
%%
%\begin{equation}
%A = N \left(\frac{\lambda}{2}\right)^2
%\end{equation}
%%
and Equation~\ref{ePG3} can be expressed as:
\begin{equation}
L_{RIS}^{-1} = \left(\frac{A}{4\pi r_i r_s}\right)^2  \left(-\hat{\bf r}_i\cdot\hat{\bf n}\right)^{2q_0} \left(+\hat{\bf r}_s\cdot\hat{\bf n}\right)^{2q_0} \epsilon_p
\label{ePG4}
\end{equation}
Equation~\ref{ePG4} indicates that path loss in the far case depends only on the physical area of the RIS, and not at all on frequency.
This is actually the expected result, as is demonstrated in the next section.

%-----------------------------------------------------------------------------
\subsection{Consistency with Plate Scattering Theory} 
%-----------------------------------------------------------------------------

Electromagnetic plate scattering theory is noted in \cite{Ozdogan+2_1910} as a possible starting point for RIS scattering models.
In this section it is shown that the connection to plate scattering theory actually follows naturally from the model developed so far, and does not need to be introduced as a postulate. 
%That is, under typical conditions the array-based model derived in previous sections yields results in which the RIS can be modeled as a passive flat plate scatterer.
To see this, % consider the far case Equation~\ref{ePG4} 
imagine that an\ RIS in the far case is replaced by a flat perfectly-conducting plate of area $A$, and let us assume monostatic geometry; i.e., 
$\hat{\bf r}_i = -\hat{\bf n}$ and
$\hat{\bf r}_s = +\hat{\bf n}$.
In this case, the radar range equation (see e.g.\ \cite{ST13}, Sec.~4.6) is:
\begin{equation}
% P_R = P_T G_T G_R \frac{\lambda^2 \sigma}{\left(4\pi\right)^3 r_i^2 r_s^2}
P_R = P_T G_T G_R \lambda^2 \sigma / \left(4\pi\right)^3 r_i^2 r_s^2
\label{eRRE}
\end{equation}
where $\sigma$ is the broadside monostatic radar cross section of the plate, which is known to be:  
\begin{equation}
%\sigma = \frac{4\pi A^2}{\lambda^2}
\sigma = 4\pi A^2 / \lambda^2
\end{equation}
see e.g.\ \cite{H61}, Sec.~3.7. 
Recasting Equation~\ref{eRRE} in the form of the Friis transmission equation, the path loss $L_{plate}$ in this scenario is given by:
\begin{equation}
% L_{plate}^{-1} = \left(\frac{A}{4\pi r_i r_s}\right)^2  
L_{plate}^{-1} = \left(A / 4\pi r_i r_s \right)^2  
\end{equation}
Now note that Equation~\ref{ePG4} gives precisely this result for $\hat{\bf r}_i = -\hat{\bf n}$, $\hat{\bf r}_s = +\hat{\bf n}$, and $\epsilon_p=1$.
Therefore Equation~\ref{ePG4} is consistent with electromagnetic plate scattering theory. 

Since the plate scattering model is both physically-rigorous and simple to compute, it serves as a useful benchmark for path loss in channels employing a RIS in the far case. 
However it should be noted that connection between plate scattering and RIS scattering is not universal. The equivalence demonstrated here is attributable to the use of the proposed $q=q_0$ element pattern model % proposed in Section~\ref{sGEPM}, 
with $\lambda/2$ element spacing.  While various other combinations of element pattern and spacing can yield the same equivalence, it is \emph{not} true that \emph{any} combination of element pattern and spacing will yield this equivalence.  
%What \emph{is} universal, however, is the $r_i^2 r_s^2$ range dependence of path loss in the far case for any passive scatterer, and thus also for any RIS.

%-----------------------------------------------------------------------------
\subsection{Comparison to Specular Reflection} 
%-----------------------------------------------------------------------------
\label{ssSpecular}

Specular reflection is the component of scattering from an electrically-large smooth surface which is distinct from diffraction originating from the edges of the surface.  If diffraction in the direction of the receiver is negligible, then, from the perspective of the receiver, the scattering from the surface is well-described as specular reflection.  It was noted from the previous section that the scattering from a RIS in the far case cannot be interpreted as specular reflection alone, since path loss is clearly seen to be dependent on the size of the RIS.
 
However, specular reflection \emph{is} commonly found to be an appropriate model for scattering from terrain, buildings, and other electrically-large structures encountered in the analysis of terrestrial wireless communications systems.  When this is the case, it is merely because diffraction is either negligible or not specifically of interest.  This begs the question:  When, if ever, is it appropriate to interpret far case RIS scattering as specular reflection?  
The short answer is ``never,'' as we shall now demonstrate.
A second finding from this analysis will be a simple guideline for choosing the size of a RIS in the far case.

Consider a RIS which is oriented such that Snell's law of reflection (i.e., angle of reflection equals angle of incidence)
%, measured with respect to locations $T$ and $R$ 
is satisfied at the RIS.
Next, imagine that the RIS is replaced by an infinitely-large flat conducting plate which lies in the plane previously occupied by the RIS. 
Since the plate is infinite, there are no edges and therefore the scattering from the plate is pure specular reflection.
Since the plate is flat, the phasefront curvature of the reflected wave at the point of reflection is equal to the phasefront curvature of the incident wave at the point of reflection, and the rate at which spatial power density decreases is the same after reflection as it was before reflection.  
Subsequently the path loss $L_S$ in this case is simply
\begin{equation}
% L_S = \left[ \frac{4\pi \left(r_i+r_s\right)}{\lambda}\right]^2 
L_S = \left[ 4\pi \left(r_i+r_s\right) / \lambda \right]^2 
\label{eLE3}
\end{equation}
i.e., the path loss is equal to that of a free-space path of length $r_i+r_s$.
Subsequently the path loss for the far case RIS channel relative to that of the specular reflection channel is:
\begin{equation}
\frac{L_S}{L_{RIS}} = \left(\frac{r_i+r_s}{r_i r_s} \cdot \frac{A}{\lambda} \right)^2  \left(-\hat{\bf r}_i\cdot\hat{\bf n}\right)^{2q_0} \left(+\hat{\bf r}_s\cdot\hat{\bf n }\right)^{2q_0} \epsilon_p
\label{eR1}
\end{equation}
Note that the ratio of the path losses is dependent on both the physical area of the RIS and frequency.
However, it is also now apparent that $L_S$ can be less than, equal to, or greater than $L_{RIS}$. 
To better understand the situation, it  is convenient to define an ``effective focal length'' $f_e$ as follows:
\begin{equation}
\frac{1}{f_e} = \frac{1}{r_i} + \frac{1}{r_s} = \frac{r_i+r_s}{r_i r_s}
\end{equation}
This expression is the known in the optics literature as the ``thin lens equation;''  
however, the reason for defining $f_e$ here is simply brevity and convenience.  
For example, when $r_i$ and $r_s$ are equal, $f_e=r_i/2=r_s/2$.
Also, when $r_i \gg r_s$, $f_e \approx r_s$; similarly when $r_i \ll r_s$, $f_e \approx r_i$.
Thus, $f_e$ generally ranges between the lesser of $r_i$ and $r_s$ down to about half the lesser value.
Using this concept,
Equation~\ref{eR1} simplifies to:
\begin{equation}
\frac{L_S}{L_{RIS}} = \left(\frac{A}{f_e\lambda}\right)^2 \left(-\hat{\bf r}_i\cdot\hat{\bf n}\right)^{2q_0} \left(+\hat{\bf r}_s\cdot\hat{\bf n}\right)^{2q_0} \epsilon_p
\label{eR2}
\end{equation}
Now we may determine the RIS size for which the path loss of these two channels is equal; i.e, $L_{RIS}/L_S=1$.  One finds:
\begin{equation}
% A = \frac{f_e \lambda}{ \sqrt{ \left(-\hat{\bf r}_i\cdot\hat{\bf n}\right)^{2q_0} \left(+\hat{\bf r}_s\cdot\hat{\bf n}\right)^{2q_0} \epsilon_p} } %~~~ \mbox{for $L_S=L_{RIS}$}
A = f_e \lambda \left[ \left(-\hat{\bf r}_i\cdot\hat{\bf n}\right)^{2q_0} \left(+\hat{\bf r}_s\cdot\hat{\bf n}\right)^{2q_0} \epsilon_p \right]^{-1/2}  %~~~ \mbox{for $L_S=L_{RIS}$}
\label{eR3}
\end{equation}

Table~\ref{tSQ1} 
\begin{table}
\caption{Side length of a square RIS for path loss equal to that of the specular reflection channel (physical dimension).}
\begin{center}
\begin{tabular}{|r||rr|rr|}
\hline
       & ``minimum'' & & ``typical'' & \\ 
Freq.  & $f_e=0.1$~km & $1$~km & $f_e=0.1$~km & $1$~km \\
\hline
 0.8~GHz &  6.1~m & 19.4~m &  8.9~m & 28.8~m \\
 1.9~GHz &  4.0~m & 12.6~m &  5.8~m & 18.2~m \\
 2.4~GHz &  3.5~m & 11.2~m &  5.1~m & 16.2~m \\
 5.8~GHz &  2.3~m &  7.2~m &  3.3~m & 10.4~m \\
28.0~GHz &  1.0~m &  3.3~m &  1.5~m &  4.7~m \\
60.0~GHz &  0.7~m &  2.2~m &  1.0~m &  3.2~m \\
\hline
\end{tabular}
\end{center}
\label{tSQ1}
\end{table}
\begin{table}
\caption{Side length of a square RIS required for path loss equal to the specular reflection channel (wavelengths).}
\begin{center}
\begin{tabular}{|r||rr|rr|}
\hline
       & ``minimum'' & & ``typical'' & \\ 
Freq.  & $f_e=0.1$~km & $1$~km & $f_e=0.1$~km & $1$~km \\
\hline
 0.8~GHz &  $16.3~\lambda$ &  $51.6~\lambda$ &  $23.7~\lambda$ & $ 74.8~\lambda$ \\
 1.9~GHz &  $25.2~\lambda$ &  $79.6~\lambda$ &  $36.5~\lambda$ & $115.3~\lambda$ \\
 2.4~GHz &  $28.3~\lambda$ &  $89.4~\lambda$ &  $41.0~\lambda$ & $129.6~\lambda$ \\
 5.8~GHz &  $44.0~\lambda$ & $139.0~\lambda$ &  $63.7~\lambda$ & $201.5~\lambda$ \\
28.0~GHz &  $96.6~\lambda$ & $305.5~\lambda$ & $140.0~\lambda$ & $442.7~\lambda$ \\
60.0~GHz & $141.4~\lambda$ & $447.2~\lambda$ & $204.9~\lambda$ & $648.0~\lambda$ \\
\hline
\end{tabular}
\end{center}
\label{tSQ2}
\end{table}
shows examples of RISs meeting this criterion. 
Two cases are considered:
``minimum,'' in which $\epsilon_p=1$ and $\hat{\bf r}_i\cdot\hat{\bf n}=1$ (i.e., broadside incidence), yielding minimum $A$; and
``typical,'' in which $\epsilon_p=0.5$ (a common value for patch-type antennas \cite{ST13}) and $-\hat{\bf r}_i\cdot\hat{\bf n}=\hat{\bf r}_s\cdot\hat{\bf n}=0.5$.
The ``typical'' case corresponds to incidence and scattering $60^{\circ}$ off broadside with realistic efficiency, representing a practical RIS in a disadvantaged geometry.
The effective focal length $f_e=0.1$~km could be a scenario in which either $r_i$ or $r_s$ is $\approx 0.1$~km with the other distance being much greater, $r_i=r_s=0.2$~km, or any number of intermediate scenarios.
Similarly, $f_e=1$~km could be a scenario in which either $r_i$ or $r_s$ is $\approx 1$~km with the other distance being much greater, $r_i=r_s=2$~km, or any number of intermediate scenarios.
Table~\ref{tSQ1} indicates that $L_{RIS}=L_S$ for RIS side-lengths ranging from meters to 10s of meters, depending on frequency and effective focal length. % in the expected way.

Table~\ref{tSQ2} shows precisely the same result, except now expressed in electrical length; i.e, units of wavelength.
Note that side-lengths ranging from $10$'s to $100$'s of wavelengths are required, as one might expect.
However, additional increases in side-length do \emph{not} yield performance comparable to specular reflection; to the contrary, the RIS outperforms specular reflection by increasing margins as the electrical size is increased.  This is simply because the RIS focuses the scattered field in order to minimize path loss, whereas as the infinite conducting plate cannot.
%\footnote{It is worth noting the same result would be obtained if the RIS were replaced by a passive flat plate of equal area, but only in the monostatic broadside case since only in that case would the scattering from the plate be ``focused.''}  
With this in mind, note also that a RIS could be used to accurately reproduce specular reflection, but the RIS would need to be both electrically-large \emph{and} configured to preserve the rate of change of phasefront curvature.  
%While this is useful in certain applications such as RIS ``broadcasting,'' it 
This is not a strategy which minimizes path loss in SISO channels. 

Note that the \emph{electrical} sizes of $A$ indicated in Table~\ref{tSQ2} increase with frequency.
In particular, note that the electrical size of $A$ increases in proportion to $\lambda^{-1/2}$, resulting in almost an order of magnitude increase as frequency increases from 0.8~GHz to 60~GHz.
  
Finally, it is noted that \cite{NJSP20} reports that the minimum RIS size that achieves path loss equal to the path loss for unobstructed direct line-of-sight between transmitter and receiver 
%(as opposed to path loss for the specular reflection channel having the same link geometry) 
is $A=r_i r_s\lambda/r_d$, where $r_d$ is the straight-line distance from transmitter to receiver.  This is simply Equation~\ref{eR3} for the special case of perfectly-efficient isotropic elements (so that the third factor in the expression becomes 1) and with $r_i+r_s$ replaced by $r_d$. 
  
%Summarizing, the notion that $A \gtrsim \mathcal{O}(100\lambda^2)$ enables scattering that is ``specular'', either literally or in terms of exhibiting the same range dependence of path loss, is incorrect in the far case.  The correct value is given by Equation~\ref{eR3}.

%%%%%%%%%%%%%%%%%%%%%%%%%%%%%%%%%%%%%%%%%%%%%%%%%%%%%%%%%%%%%%%%%%%%%%%%%%%%%%
\section{General Case}
%%%%%%%%%%%%%%%%%%%%%%%%%%%%%%%%%%%%%%%%%%%%%%%%%%%%%%%%%%%%%%%%%%%%%%%%%%%%%%
\label{sRSNF}

\begin{figure} %[!t]
\begin{center}
\includegraphics[width=0.9\columnwidth]{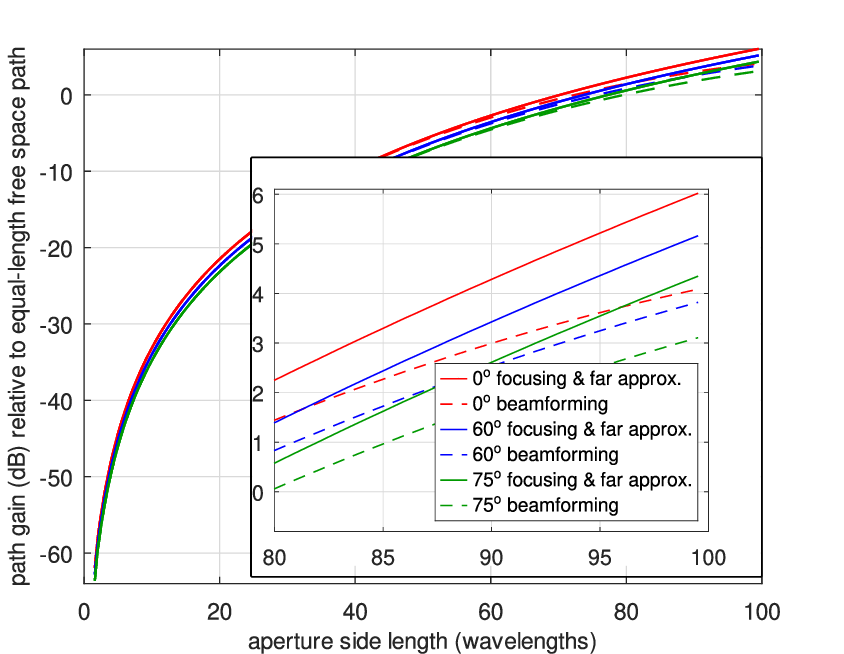}
\end{center}
\caption{
Path gain relative to the equal-length free space channel for $r_i=r_s=10^4\lambda$ and various $\psi_s$.
Inset shows detail in upper right corner.
}
\label{fPG10000}
\end{figure}

\begin{figure} %[!t]
\begin{center}
\includegraphics[width=0.9\columnwidth]{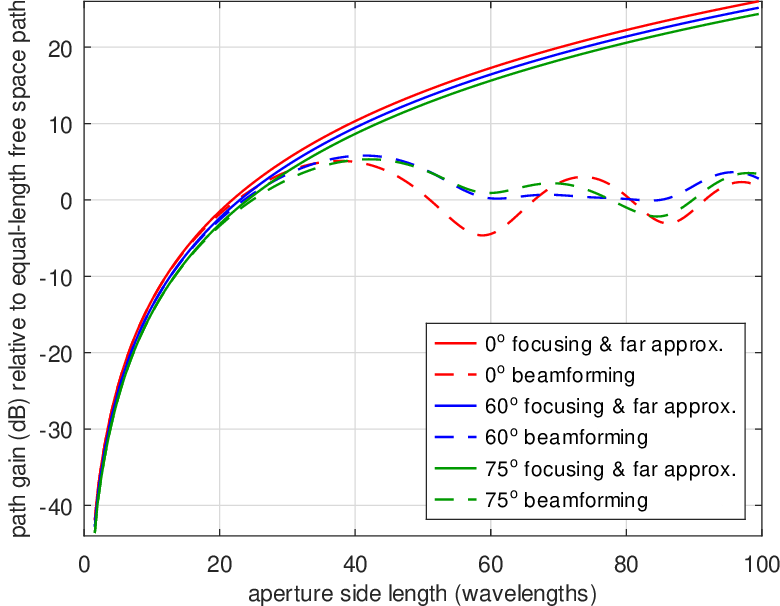}
\end{center}
\caption{
Path gain relative to the equal-length free space channel for $r_i=r_s=10^3\lambda$ and various $\psi_s$.
}
\label{fPG1000}
\end{figure}

\begin{figure} %[!t]
\begin{center}
\includegraphics[width=0.9\columnwidth]{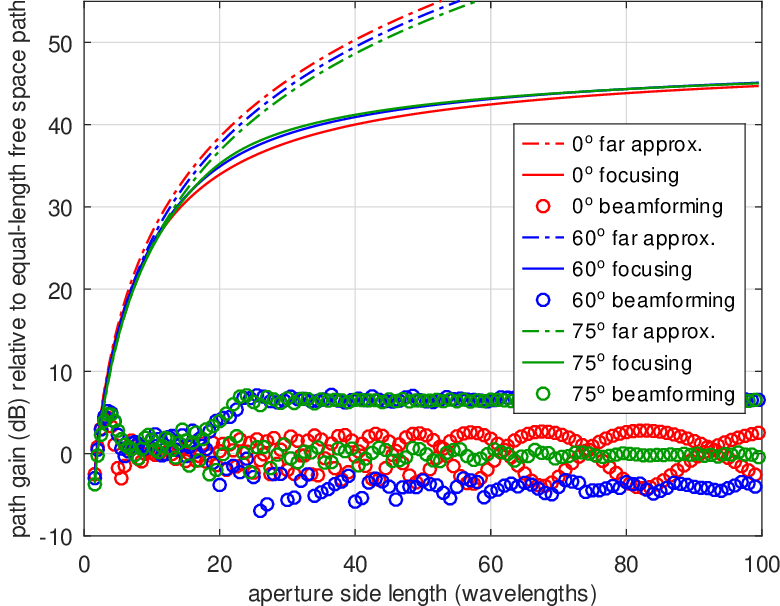}
\end{center}
\caption{
Path gain relative to the equal-length free space channel for $r_i=r_s=10\lambda$ and various $\psi_s$.
%Curves are shown for $\psi_s=0^{\circ}$ (\emph{red}), $60^{\circ}$ (\emph{blue}), and $75^{\circ}$ (\emph{green}).
Beamforming results are shown as unconnected data points to make clear the rapid variation even when varying the aperture size by the minimum amount; i.e., by one additional row and column of elements per sample. 
}
\label{fPG10}
\end{figure}

When the ``far'' criteria defined at the beginning of Section~\ref{sFFC} are not met, one is forced to backtrack to Equation~\ref{ePG2}.
In this case, the conclusions in the far case do not apply and analogous conclusions for the general case are not obvious.
For this reason, the following numerical study is conducted.
This study assumes a planar square RIS which is again modeled as $N$ elements with pattern $q=q_0$ which are uniformly distributed with $\lambda/2$ spacing.
The transmitter is broadside to the array while the angle $\psi_s$ to the receiver is varied from $0$ (RIS broadside) to $60^{\circ}$ and then $75^{\circ}$.  
The distances $r_i$ and $r_s$ are held equal and varied to generate the results shown in Figures~\ref{fPG10000}, \ref{fPG1000}, and \ref{fPG10}.

%%-----------------------------------------------------------------------------
%\subsection{Focusing vs.\ Beamforming}

In the general case, the path loss minimization criterion $b_n=e^{-j\phi_n}$ is referred to as ``focusing''.  
A different option is to set the phase of $b_n$ so as to minimize path loss with respect to the directions (only) of the transmitter and receiver, which requires only that $\hat{\bf r}_i$ and $\hat{\bf r}_s$ (and not the distances $r_i$ or $r_s$) be known.  
Specifically:
\begin{equation}
b_n = e^{-j(2\pi/\lambda){\bf p}_n\cdot\hat{\bf r}_i}  e^{+j(2\pi/\lambda){\bf p}_n\cdot\hat{\bf r}_s}
\end{equation}
where ${\bf p}_n$ is the vector from a common reference point (e.g., the origin) to the location of element $n$.
For the purposes of this paper, this is referred to as ``beamforming,'' and can be alternatively interpreted as collimation, or focusing at infinity.  Beamforming is of practical interest, despite the non-optimum path loss, because in practice it may be easier to determine directions than positions.  
%This distinction appears in previous work \cite{x,y} but is not clearly identified as a choice between focusing (requiring either precise location information or precise channel state estimation) and beamforming (requiring only directions).
The distinction between focusing and beamforming was not necessary in the far case, since under the far case approximations they are equivalent.
However, when the RIS is closer (and as correctly noted in \cite{BS20} in particular), the difference in performance can be large, as we shall now see.

%%-----------------------------------------------------------------------------
%\subsection{Results}

Fig.~\ref{fPG10000} shows path gain $L_{RIS}^{-1}$ from Equation~\ref{ePG2}, computed for $r_i=r_s=10^4\lambda$.
The result is normalized to the path gain $L_S^{-1}$ (Equation~\ref{eLE3}) for a free space (i.e., no RIS) channel having path length $r_i+r_s$.
Thus, 0~dB on the vertical axis indicates that the RIS-enabled channel exhibits path gain equal to that of the equal-length free space path.   
Separate curves are shown for focusing, beamforming, and the far approximation (Equation~\ref{ePG4}).
The close agreement (especially between focussing and the far approximation) indicates that this scenario meets the far approximation criteria.
%Therefore the far approximation is very good for RIS apertures up to at least $100\lambda \times 100\lambda$ ($N=10^4$ elements) at a distance of $10^4\lambda$.
Also, note that an aperture side-length of at least $70\lambda$ is required to meet the 0~dB condition under these conditions.

Fig.~\ref{fPG1000} shows the results of the same experiment performed for $r_i=r_s=10^3\lambda$; i.e., the same RIS but one order of magnitude closer.  
Again, separate curves are shown for focusing, beamforming, and the far approximation.
In this case, focusing and the far approximation remain too close to distinguish.
Therefore the far approximation is very good for RIS apertures up to at least $100\lambda \times 100\lambda$ ($N=10^4$ elements) for distances $\gtrsim 10^3\lambda$.
The performance of beamforming, on the other hand, is seen to be limited to within a few dB of 0~dB for aperture side-lengths greater than about $20\lambda$.
In this sense, a RIS operating in beamforming mode might be said to be exhibiting performance comparable to that of specular reflection.
This is a particularly useful insight: If beamforming is to be used, then there is a maximum useful size (given by Equation~\ref{eR3}) for the RIS, and further increases in size will not significantly reduce path loss. 
Essentially the same observation is made in \cite{BS20} using a somewhat different method; in fact Equation~55 in \cite{BS20} is Equation~\ref{eR3} for the special case of a perfectly-efficient RIS and broadside monostatic link geometry.
If, on the other hand, focusing is used, then further increases in size \emph{do} significantly reduce path loss.

Fig.~\ref{fPG10} shows the results of the same experiment performed for $r_i=r_s=10\lambda$.
In this scenario, the RIS side-length becomes much larger than the distances to the transmitter and receiver, and so the far approximation is expected to fail.
This is observed to be the case, with the results for focusing and the far approximation diverging for side-lengths greater than about $r_i$ (or $r_s$).
Focusing approaches 45~dB normalized path gain for $100\lambda$ side-length, compared to $0\pm 6$~dB (independent of side length) for beamforming.   
Thus -- as expected -- focusing is capable of enormous gain over beamforming when path distances are less than the aperture side-length.

The model developed in this paper does not account for variation in the per-element incident and scattered polarizations or variation in transmitter and receiver antenna gains over the RIS.
It is noted in \cite{BS20} that this may introduce significant error when $r_i$ or $r_s$ are small compared to the size of the RIS, as is the case in the scenario addressed in Fig.~\ref{fPG10}. 
While the results shown in Fig.~\ref{fPG10} exhibit the expected behavior, they are nevertheless less accurate and should be used with caution.

%%%%%%%%%%%%%%%%%%%%%%%%%%%%%%%%%%%%%%%%%%%%%%%%%%%%%%%%%%%%%%%%%%%%%%%%%%%%%%
\section{Conclusions}
%%%%%%%%%%%%%%%%%%%%%%%%%%%%%%%%%%%%%%%%%%%%%%%%%%%%%%%%%%%%%%%%%%%%%%%%%%%%%%
\label{sConc}

This paper presented a simple physical model for RIS scattering and employed the model to develop expressions for path loss.
To avoid the need for implementation-specific RIS design parameters, a physically-motivated benchmark element pattern model was proposed.  This pattern has the useful feature that the sum of the effective apertures of the elements is equal to the physical area of the RIS when the element spacing is $\lambda/2$.  
When a RIS with the benchmark element pattern and $\lambda/2$ spacing is analyzed using this model, the results are shown to be consistent with both electromagnetic plate scattering theory and prior work.
Other RISs can be represented by varying the element parameter $q$ and the element spacing.
The principal shortcoming of the model is reduced accuracy when either transmitter or receiver is very close to the RIS, since the per-element magnitude, phase, and polarization response can vary significantly across the array in this case, whereas the model as presented here accounts only for variation in the magnitude and phase, but not polarization. 
The fully polarimetric models of \cite{BS20}, \cite{NJSP20}, and \cite{GSK20} are not subject to this limitation.
However, it would be straightforward to incorporate polarization into the model presented in this paper as well.

\end{document}